\documentclass[twocolumn, prx, reprint, amsfonts, amsmath, amssymb, superscriptaddress, longbibliography, floatfix,aps]{revtex4-2}
\usepackage[T1]{fontenc}
\usepackage{graphicx}
\usepackage[dvipsnames]{xcolor}
\usepackage[colorlinks,linkcolor=Blue,citecolor=Blue]{hyperref}
\usepackage{braket}

\begin{document}

\title{The Floquet Fermi Liquid}

\author{Li-kun Shi}
\thanks{These authors contributed equally to this work.}
\affiliation{Institut f{\"u}r Theoretische Physik, Universit{\"a}t Leipzig, Br{\"u}derstra{\ss}e 16, 04103, Leipzig, Germany}

\author{Oles Matsyshyn}
\thanks{These authors contributed equally to this work.}
\affiliation{Division of Physics and Applied Physics, School of Physical and Mathematical Sciences, Nanyang Technological University, Singapore 637371, Republic of Singapore}

\author{Justin C. W. Song}
\affiliation{Division of Physics and Applied Physics, School of Physical and Mathematical Sciences, Nanyang Technological University, Singapore 637371, Republic of Singapore}

\author{Inti Sodemann Villadiego}
\email[]{sodemann@itp.uni-leipzig.de}
\affiliation{Institut f{\"u}r Theoretische Physik, Universit{\"a}t Leipzig, Br{\"u}derstra{\ss}e 16, 04103, Leipzig, Germany}

\date{\today}

\begin{abstract}
We demonstrate the existence of a non-equilibrium ``Floquet Fermi Liquid'' state arising in partially filled Floquet Bloch bands weakly coupled to ideal fermionic baths, which possess a collection of ``Floquet Fermi surfaces'' enclosed inside each other, resembling matryoshka dolls. We elucidate several properties of these states, including their quantum oscillations under magnetic fields which feature slow beating patterns of their amplitude reflecting the different areas of the Floquet Fermi surfaces, consistent with those observed in microwave induced resistance oscillation experiments. We also investigate their specific heat and thermodynamic density of states and demonstrate how by controlling properties of the drive, such as its frequency, one can tune some of the Floquet Fermi surfaces towards non-equilibrium van-Hove singularities without changing the electron density.
\end{abstract}

\maketitle

\textit{\color{blue}Introduction.} This study addresses the fate of Fermi liquids and their Fermi surfaces when they are driven far away from equilibrium by periodic time-dependent perturbations. To address this we will revisit a more general problem that has attracted recent attention \cite{seetharam2015controlled,dehghani2015out,iadecola2015occupation,bilitewski2015scattering,genske2015floquet,esin2018quantized,uhrig2019positivity,park2022steady,matsyshyn2023fermi,kumari2023josephson}: how should Floquet states be occupied by fermions? To answer this question it is important to consider a system in contact with a bath, because periodically driven closed systems that are thermalizing tend to have trivial infinite temperature steady states~\cite{d2014long,lazarides2014equilibrium,ponte2015periodically}, and those that are not thermalizing tend to retain memory of initial conditions~\cite{lazarides2014periodic,lazarides2014equilibrium,lazarides2015fate,khemani2016phase,else2016floquet,else2017prethermal}, making their steady states not unique. We will consider an ``all fermion'' setting, where the system and the bath are both comprised only of fermions.

\begin{figure}[t]
\includegraphics[width=0.48\textwidth]{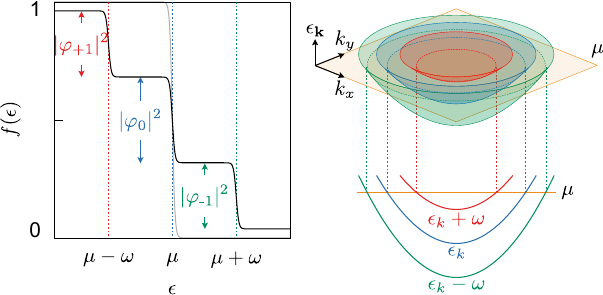}
\caption{Fermi Dirac Staircase occupation of Floquet states (left), from Eq.\eqref{CleanLimit-Periodic-DensityMatrix}, and its associated Floquet Fermi surfaces (right) from Eq.\eqref{FFSdef}.}
\label{fig-schematic}
\end{figure}

Within such setting, we have a found a remarkable answer to this question: a non-equilibrium steady state with a sizable energy density difference relative to the ground state but which retains its quantum nature, which we call the Floquet Fermi Liquid. Unlike its equilibrium counterpart where states are occupied according to the Fermi Dirac distribution, the Floquet Fermi Liquid features a staircase-shaped occupation of the Floquet band with multiple jumps that evolve into sharp discontinuities at zero temperature giving rise to a collection of enclosed Floquet Fermi Surfaces (See Fig.\ref{fig-schematic}). We will investigate the fingerprints left by Floquet Fermi Surfaces in various observables, such as the appearance of a slow beating of the quantum oscillations amplitude, as well as the density of states and the specific heat.

\medskip

\noindent \textit{\color{blue}Fermi Dirac Staircase Periodic Gibbs Ensemble.}
Consider a model of non-interacting fermions in contact with a fermionic bath, with a single particle Hamiltonian of the system plus bath of the form:
\begin{align}\begin{aligned}
H (t) = \begin{bmatrix}
H_S (t) & H_{SB} \\
H_{BS} & H_B
\end{bmatrix} .
\label{Full-Hamiltonian}
\end{aligned}\end{align}
The system can be viewed as a tight-binding model, where each site can tunnel (via $H_{SB}$) to a collection of bath sites that are a set of independent energy levels (described by $H_{B}$). This is a ``grand-canonical'' setting where the energy and particle number of the system can fluctuate.  We assume the bath to be ``featureless'', namely with energy-independent density of states and tunneling amplitudes over a band-width that is much larger than the system's, as it is frequently assumed ~\cite{jauho1994time,johnsen1999quasienergy,kohler2005driven,oka2009photovoltaic,kamenev2011field,kitagawa2011transport,morimoto2016topological,matsyshyn2021rabi,park2022thermal,shi2023berry,matsyshyn2023fermi}.

Crucially, the system Hamiltonian, $H_S (t)$, can be time-dependent, allowing us to drive it away from thermal equilibrium with the bath. By assuming that the bath is prepared in a thermal ensemble in a distant past with a Fermi Dirac distribution, $f_0 (\epsilon)=1/[1+e^{\beta(\epsilon-\mu)}]$
with inverse temperature $\beta$ and chemical potential $\mu$, one can rigorously show (see Appendix \ref{appendixA1}), that at late times the system approaches a unique steady state, with its exact one-body density-matrix given by:
\begin{align}\begin{aligned}
& \rho_S (t) = \Gamma 
\int_{-\infty}^{+\infty} \frac{ \text{d} \epsilon }{\pi}
f_0 (\epsilon) U_\Gamma (t, \epsilon) U_\Gamma^\dagger (t, \epsilon),
\\
& U_\Gamma (t, \epsilon) = \int_{-\infty}^{t} \text{d} t' 
e^{\Gamma (t' - t) - i \epsilon  t' } U_S (t, t') ,
\label{Full-DensityMatrix}
\end{aligned}\end{align}
where $U_S (t,t')$ satisfying $i \partial_t U_S (t,t') = H_S (t) U_S (t,t') $ is the unitary evolution operator of the isolated system and $\Gamma = \lambda^2 \nu_0/ 2$ is the particle tunneling rate into the bath, which parametrizes the strength of system-bath coupling. Equation \eqref{Full-DensityMatrix} generalizes Eq.(41) of Ref.~\cite{matsyshyn2023fermi} to arbitrary off-diagonal time-dependent system Hamiltonians. Now by assuming that the drive is periodic, $H(t)=H(t+T)$, and the coupling to the bath is infinitesimal, $\Gamma \to 0$, so that it would act as an ``ideal'' bath in equilibrium, then Eq.~(\ref{Full-DensityMatrix}) reduces to (see appendix \ref{appendixA2}):

\begin{align}\begin{aligned}
& \lim_{\Gamma \to 0} \rho_S (t) = \sum_a f_a \ket{\psi_a^F (t)} \bra{\psi_a^F (t)},
\\
& f_a = \sum_{l =-\infty}^{+\infty} \big| \varphi_{a,l} \big|^2 f_0 (\epsilon_a^F + l \Omega) ,
\label{CleanLimit-Periodic-DensityMatrix}
\end{aligned}\end{align}
Here $\ket{\psi_a^F (t)} $ are the complete basis of solutions of the single particle time dependent Schrodinger equation,  $\epsilon_a^F$ are their Floquet energies, $\Omega = 2\pi/T$, $\big| \varphi_{a,l} \big|^2 \equiv  \braket{\varphi_{a,l}|\varphi_{a,l}}$ with $\ket{\varphi_{a,l}}$ the $l$-th harmonic of the Floquet wave-function, related as $\ket{\psi_a^F (t) }
= \sum_l e^{- i \epsilon_a^F t - i l \Omega t} 
\ket{\varphi_{a,l}} $ \footnote{These jumps satisfy the normalization condition $\sum_l\big| \varphi_{a,l} \big|^2=1$, and are therefore constrained to satisfy $| \varphi_{a,l} \big|^2\leq 1$}.

Equation~(\ref{CleanLimit-Periodic-DensityMatrix}) is an example of a {\it periodic Gibbs ensemble} ~\cite{lazarides2014periodic,lazarides2014equilibrium,lazarides2015fate,khemani2016phase}, but in contrast to the setting of Refs.~\cite{lazarides2014periodic,lazarides2014equilibrium,lazarides2015fate,khemani2016phase} we have obtained this ensemble by coupling the system to a bath and not as a result of many-body self-thermalization. In the context of self-thermalization the occupations, $f_a$, would not be fixed but determined by initial conditions of the quasi-particles, but in our context the $f_a$ are uniquely fixed by the state of the bath. Notably the occupations, $f_a$, viewed as a function of the Floquet energy, $\epsilon_a^F$, are not given by the equilibrium Fermi-Dirac function but instead by a Fermi-Dirac Staircase (see Fig.~\ref{fig-schematic}), generalizing the results of Ref.~\cite{matsyshyn2023fermi} to off-diagonal Hamiltonians. These staircase occupations have also appeared in Eq.~(12) of Ref.~\cite{seetharam2015controlled} and Eq.(1) of Ref.~\cite{kumari2023josephson}, and in discussions of the Tien-Gordon effect \cite{nazarov2009quantum,tien1963multiphoton} in driven mesoscopic systems.

\medskip

\noindent \textit{\color{blue}The Floquet Fermi Liquid.} Let us now specialize to the case of a Floquet Bloch band. For simplicity, we take a system with a single band arising from a tight-biding model with one site per unit cell with dispersion $\epsilon({\bf k})$, and driven by a time-periodic and spatially uniform electric field with vector potential ${\bf A}(t)={\bf A}(t+T)$ so that the Hamiltonian remains diagonal in crystal momentum and is given by $\epsilon_{\bf k}  (t) \equiv \epsilon({\bf k}-{\bf A}(t))$. In this case, the density matrix is indeed time-independent and the occupations can be obtained from Eq.\eqref{CleanLimit-Periodic-DensityMatrix} by replacing $a\rightarrow {\bf k}$: 
\begin{align}\begin{aligned}
f_{\bf k} = \sum_{l} |\varphi_{{\bf k}, l}|^2 f_0 ( \epsilon_{\bf k}^F + l \Omega) .
\label{clean_p_k}
\end{aligned}\end{align}
where the Floquet energy and the harmonics of the Floquet wavefunctions are given by $\epsilon_{\bf k}^F  = \langle \epsilon_{\bf k} (t)\rangle_T$, $ \varphi_{{\bf k}, l} = \langle e^{-i \int_{0}^{t} {\text d} t' [\epsilon_{\bf k} (t') - \epsilon_{\bf k}^F - l \Omega ]}\rangle_T$, and $ \langle \cdots \rangle_T  = \int_{0}^T  (\cdots){{\text d} t}/{T}$ denotes the time average over one period. $f_{{\bf k}}$ in Eq.(\ref{clean_p_k}) describes the occupation of canonical crystal momentum ${\bf k}$, which is related to the physical gauge invariant crystal momentum via ${\bf k}_{\rm phys} = {\bf k} -{\bf A}(t)$. Since the occupation of canonical momenta is time-independent, the occupation of physical momenta oscillates as ${\bf A}(t)$. Therefore, the occupation develops a collection of sharp steps at several surfaces in crystal momentum that are enclosed inside each other and are given by:
\begin{align}\begin{aligned}
\epsilon_{\bf k}^F  = \mu - l \Omega, 
\qquad 
l \in \mathbb{Z}.
\label{FFSdef}
\end{aligned}\end{align}

\noindent We will refer to these surfaces as the Floquet Fermi Surfaces (FFS) and the corresponding non-equilibrium steady state as the Floquet Fermi Liquid (FFL). Notice that the height of the jump at the $l$-th FFS, given by $| \varphi_{{\bf k},l} |^2 $, is in general a function of the momentum within a given FFS.

\medskip

\noindent \textit{\color{blue}Quantum Oscillations of the FFL.} As we have seen, a periodically driven system of fermions in contact with a fermionic bath approaches a non-trivial FFL steady state with a collection of enclosed FFS's. We would like to investigate how these FFS's manifest directly through observable properties \footnote{We note that a few previous studies have indeed indirectly dealt with the FFL through its manifestation in properties such as RKKY interactions \cite{asmar2021floquet}, susceptibility functions \cite{ono2019nonequilibrium}, and staircase occupations have also been discussed in driven Luttinger liquids~\cite{graf2010parametric,kagan2009parametric,bukov2012parametric}.}. 

Systems with a Fermi surface display characteristic quantum oscillations of many observables in the presence of applied magnetic fields, with a periodicity of the form $\sim \cos(S/B)$, where $S$ is the area of the Fermi surface. As we will show, the FFSs give rise to a sum quantum oscillations with different frequencies $\sim \cos(S_l / B)$, where $S_l$ is the area of the $l$-th FFS. We will show that in the regime where the cyclotron energy is smaller than the driving frequency this will lead to a slow beating of the amplitude of quantum oscillations, which, remarkably, has the same period measured in two-dimensional electron systems in the regime where microwave induced resistance oscillations (MIRO) coexist with the Shubnikov–de Haas (SdH) oscillations~\cite{shi2015shubnikov,dmitriev2012nonequilibrium}, suggesting that the Floquet Fermi liquid has indeed already been achieved in these experiments.

To illustrate this, we consider parabolic fermions, coupled simultaneously to a uniform magnetic field, ${\bf B} = \nabla \times {\bf A}_{0}({\bf r})$, and a time-dependent electric field, ${\bf E}(t) = -\partial_t {\bf A}(t)= {\bf E} e^{-i\Omega t}+c.c.$, with Hamiltonian: $H_S(t) = [{\bf k}-{\bf A}_{0}({\bf r})-{\bf A}(t)]^2/(2m)$. The solutions of the time dependent Schrödinger equation for this Hamiltonian are time-dependent Landau levels wave-functions, ${\ket{\psi^F_N(t)}}$, with Floquet energy $\epsilon^F_{N}$ and labeled by a principal cyclotron index $N=0,1,2..$, and with a guiding-center degeneracy $N_\phi$ (see Appendix \ref{C2}). By replacing $a\rightarrow N$ in the formula for the steady state from Eq.(\ref{CleanLimit-Periodic-DensityMatrix}), we can compute various observables of the system. Here we will focus on the oscillations of an effective Floquet free energy defined as follows:
\begin{align}\begin{aligned}
\beta G 
\equiv -N_\phi
\sum_{N,l} 
|\varphi_{N,l}|^2 
\log \Big[1 + e^{-\beta (\epsilon^F_{N}+l\Omega-\mu)}\Big].
\label{Geff}
\end{aligned}\end{align}
This free energy approaches the equilibrium free energy in the limit of static Hamiltonians (${\bf A}(t)\rightarrow 0$): $\beta G_{\rm eff}\rightarrow -\ln Z(\beta,\mu)$, where $Z(\beta,\mu)$ is the Grand-canonical  partition function in the absence of drive \footnote{Conceptually, the effective Floquet free energy in Eq.\eqref{Geff}, can be viewed as a sum of the effective free energies of each Floquet quasi-energy state, $\epsilon^F_{N}+l\Omega$, weighed by the amplitude of the corresponding Harmonic of the Floquet wavefunction  $|\varphi_{N,l}|^2$, and thus it is a mathematically natural extension of the relevant equilibrium free energy in a grand-canonical ensemble to a Floquet system, although its physical significance is not as transparent. We have utilized this free energy for conceptual illustration in the main text because it displays much simpler oscillations than other conceptually more natural quantities such as the energy averaged over one period. Nevertheless, the theory of the oscillations of such time averaged energy are presented in Appendix  \ref{appendixC}.}. In the absence of magnetic fields, and to second order in driving electric fields, only the Floquet bands shifted in energy by $\pm \hbar\Omega$ contribute. In a magnetic field to second order in electric fields, we would have two Floquet copies of the Landau level spectrum (see Fig.[\ref{fig-osc}]), since higher Floquet harmonics have weights with higher powers of electric field (see Appendix \ref{appendixB}). We will thus compute this free energy to second order in the electric fields. By performing a similar analysis to the equilibrium calculation \cite{shoenberg2009magnetic}, we have found the oscillating part of the Floquet free energy in the limit where many Landau levels are occupied $\mu \gg \hbar \omega_c$ is (see Appendices \ref{C2} and \ref{appendixC} for details):

\begin{figure}[t]
\includegraphics[width=0.4\textwidth]{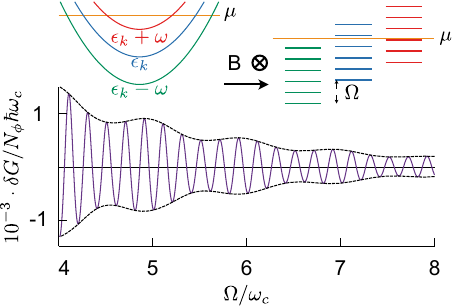}
\caption{Illustration of Floquet Landau levels (top right) and free energy oscillations computed from Eq.(\ref{Gosc}), for $T = 0.05\mu, \Omega=0.2\mu, {\bf E}/(\Omega\sqrt{m\mu}) =0.05\cdot\mathbf{n_E}/\|\mathbf{n_E}\|$, $\mathbf{n_E}=(1, i/2)$.}
\label{fig-osc}
\end{figure}

\begin{align}\begin{aligned}
\frac{\delta G}{N_\phi \hbar\omega_c} \approx 
\sum_{l =\pm1,0}\sum_{k=1}^\infty G_k R_k
\Big(\delta_{l,0} + \frac{b_l \mu_l}{\hbar\omega_c}\Big)
\cos\Big( \frac{k S_l}{B}\Big),\label{Gosc}
\end{aligned}\end{align}

\noindent where $G_k=(-1)^k/(2k^2\pi^2)$, $R_k = \lambda k /\sinh(\lambda k )$ is the Lifshitz-Kosevich factor, $\lambda=2\pi^2/(\beta\hbar\omega_c)$, $S_l \approx 2m\pi(\mu - l \Omega)$ is the area of $l$-th FFS, $b_{\pm1} =-b_0/2 = 
R_+ = {|z_+|^2}/{(\omega_c-\Omega)^2}+{|z_-|^2}/{(\omega_c+\Omega)^2},$
and $|z_\pm|^2 = \omega_c(|{\bf E}|^2\pm i[{\bf E}\times{\bf E}^*]_z)/(2m\Omega^2)$. We therefore see that the FFSs give rise to additional frequencies of the quantum oscillations controlled by their effective areas, resembling a multi-band system in equilibrium, as illustrated in Fig.~(\ref{fig-osc}).

The above oscillations resemble closely those observed in two-dimensional electron systems in the regime where MIRO and SdH oscillations coexist \cite{shi2015shubnikov}, where a rich variety of non-equilibrium phenomena have been observed \cite{zudov1997microwave,zudov2001shubnikov,ye2001giant,lei2009temperature,dmitriev2011nonequilibrium}, that also have been realized for electrons in the surface of Helium \cite{MIROHelium} and more recently in graphene \cite{mönch2020observation}. To make a more direct connection with these, let us compute also the oscillations of an effective non-equilibrium thermodynamic density of states (DoS), defined as:
\begin{align}\begin{aligned}
    \nu \equiv \left(\frac{\partial n}{\partial \mu}\right)_T, \ n = \frac{1}{2 \pi l^2} \sum_{N} f_N,
\end{aligned}\end{align}
where $l^2=\hbar c/e B$ is the magnetic length, and $\mu, T$ are the chemical potential and temperature of the bath. This non-equilibrium DoS reduces to the equilibrium DoS in the absence of drive, and the oscillations of DoS tend to resemble those of resistivity in equilibrium \cite{shoenberg2009magnetic,abrikosov2017fundamentals}, making them a more relevant observable to contrast with MIRO photoconductivity measurements. The oscillatory part of the DoS (see Appendix \ref{appendixC}), can be shown to be:
\begin{align}\begin{aligned}
\delta \nu 
\approx 
\frac{2}{h\omega_c l^2}
\sum_{k=1}^\infty
(-1)^k R_k F_E 
\cos\Big(k \frac{S}{B}\Big),
\end{aligned}\end{align}
where the factor $F_E =  1 - 4R_+({\mu}/{\hbar\omega_c})\sin^2(\pi{\Omega}/{\omega_c})$ describes the oscillations of the envelope of the fast oscillations (see Fig.[\ref{fig-osc}]), imprinted by the AC drive. The frequency and phase of these envelope oscillations agrees exactly with that of photo-resistivity theories from Refs.\cite{dmitriev2011nonequilibrium,lei2009temperature}. The frequency of oscillations of the envelope also agrees with those of the photo-resistivity in MIRO experiments but not with their phase \cite{shi2015shubnikov}, for which there is no current detailed understanding, although it is expected to depend on the intensity of radiation \cite{dmitriev2011nonequilibrium}, and on details of the scattering mechanisms \cite{lei2009temperature}. Therefore, The FFL and its collection of FFSs, provides a simple overarching conceptual framework that positions MIRO as a natural non-equilibrium counterpart to conventional equilibrium quantum oscillations. We hope this picture can contribute to clarify and guide experiments in the future \footnote{Our fermionic bath is by no means a realistic approximation to the relevant relaxation mechanisms in typical 2DEGs where MIRO is observed. We believe, however, that the existence of multiple FFS is a robust phenomenon that remains in the presence of other more realistic relaxation agents, such as phonons. Some properties of the quantum oscillations of FFL's, such as the slow beating frequency of their amplitude, will be robust to the details of relaxations because they are controlled by the difference of areas of FFS's. However other aspects of the oscillations such as their amplitude and phase, will be more sensitive to the detailed relaxation mechanisms, which is an important question to address in future work.}.

\medskip

\begin{figure}[t]
\includegraphics[width=0.48\textwidth]{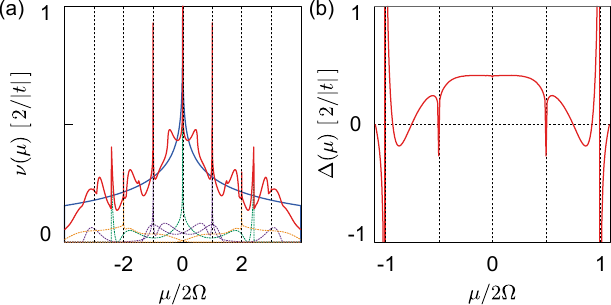}
\caption{(a) DoS for the non-driven (blue) and driven (red) square lattice model from Eq.(\ref{Density-of-states}). Dashed lines are the additional van-hove singularities from Eq.\eqref{vanHovemu}. Green, purple, and orange lines are contributions making the red-line coming from $l=0, \pm 2, \pm 4$ harmonics respectively.
(b) Correction $\Delta(\mu)$ to the equilibrium specific heat coefficient from Eq.(\ref{linear-T-coefficient}). Parameters: $\Omega = 1/4$, $A_x = A_y = 4/3$, $\phi_x = 0$, $\phi_y = \pi/2$.}
\label{fig-DOS-CV}
\end{figure}

\noindent \textit{\color{blue} DoS and non-equilibrium van-Hove singularities of the FFL.}
The thermodynamic DoS plays a central role in equilibrium and is directly measurable via capacitive measurements of compressibility~\cite{smith1986two,eisenstein1992negative,martin2008observation,feldman2012unconventional,yang2021experimental}~\footnote{There are also optical measurement techniques \cite{xia2023optical}).}. Notably, a non-interacting system with its chemical potential tuned at a van-Hove singularity, for which the DoS diverges, would generically become unstable towards broken symmetry states for weak interactions (see e.g. Ref.\cite{nandkishore2012chiral}). Here we would like to demonstrate that FFLs possess a greater degree of tunability relative to their equilibrium counterparts, because the parameters controlling the radiation, such as the frequency, can be used to tune it towards a van-Hove singularity of its non-equilibrium DoS, without the need to change the electron density. To demonstrate this we consider a single band model. Using Eq.~(\ref{clean_p_k}), the non-equilibrium DoS can be expressed as a sum of an effective DoS of each FFS:
\begin{align}\begin{aligned}
& \nu(\mu)= \lim_{T \to 0}
(  \partial n/\partial \mu )_T
=
\sum_{l} \nu_l (\mu),
\\
& \nu_l (\mu) = 
\int \frac{d^d {\bf k} }{(2 \pi)^d} 
\big| \varphi_{{\bf k}, l} \big|^2 \delta (\mu - l \Omega - \epsilon_{\bf k}^F).
\label{Density-of-states}
\end{aligned}\end{align}
Therefore, the frequency can be used to shift the effective chemical potential of $l$-th FFS as $\mu-l\Omega$. As an example, consider a 2D square lattice with nearest neighbor hopping amplitude $t$, so that in equilibrium it would have dispersion $\epsilon_{\bf k} 
= -2 t \cos(k_x) -2 t \cos(k_y)$, with a van-Hove singularity at $\mu = 0$ originating from the states near the two special momenta $(\pi,0),(0,\pi)$ (see Fig.~\ref{fig-DOS-CV}(a)). In the driven case, ${\bf A} (t) = [A_x \sin (\Omega t + \phi_x), A_y \sin (\Omega t + \phi_y) ]$, the Floquet band energy is (see Appendix \ref{appendixF}):

\begin{align}\begin{aligned}
& \epsilon_{\bf k}^F
= -2 t \big[
\cos (k_x) J_0 (A_x)
+
\cos (k_y) J_0 (A_y)
\big],
\end{aligned}\end{align}

\noindent where $J_0$ is the Bessel function of first kind. This Floquet problem retains a ${\bf k} \to - {\bf k}$ symmetry which pins the origin of van-Hove singularities of the higher order FFSs to the same two special momenta $(\pi,0),(0,\pi)$. This symmetry also leads to a vanishing of the odd Floquet wave-functions at these momenta, namely $\varphi_{(0,\pi), l} = \varphi_{(\pi,0), l} = 0$
for odd $l$ 
(see Appendix \ref{appendixF}). However, for $l$ even, the Floquet amplitudes remain finite near these points and as a result such FFSs display additional van-Hove singularities in the non-equilibrium DoS, at the following chemical potentials (see Fig.~\ref{fig-DOS-CV}(a)):

\begin{align}\begin{aligned}\label{vanHovemu}
\mu = l \Omega \pm 
\big[ 
J_0 (A_x)
-J_0 (A_y) \big], \ l \ \rm{even} .
\end{aligned}\end{align}

This model illustrates the tantalizing potential of engineering the properties of the AC drive to tune some FFSs into van-Hove singularities, even if at equilibrium there is no DoS singularity at the chemical potential. 

\noindent \textit{\color{blue} Non-equilibrium specific heat of the FFL.} In equilibrium, the low temperature  specific heat, $C_V$, in a Landau-Fermi liquid is related to the thermodynamic DoS via \cite{pines2018theory}:
\begin{align}\begin{aligned}
\lim_{T \to 0 }
\frac{C_V}{\frac{\pi^2}{3} k_B T } 
= (1 + F_0^s) 
\lim_{T \to 0}
\Big( \frac{\partial n}{\partial \mu} \Big)_T,
\end{aligned}\end{align}
where $C_V$ is the specific heat at constant volume, and $F_0^s$ is the spin symmetric Landau parameter, which is non-zero only in the presence of interactions. Therefore for non-interacting fermions, the linear in $T$ coefficient of the specific heat also measures the DoS. However, interestingly, the FFLs violate the above relation between specific heat and thermodynamic DoS even for non-interacting fermions. To show this, we begin by defining the non-equilibrium the specific heat, $C_V$, as $C_V\equiv (\partial \bar{E}/\partial T)_n$, where  $\bar{E}$ is the system energy time-averaged over one period, and the derived is computed at fixed particle density. For our model with single Floquet Band, we obtain the following relation (see Appendix \ref{appendixE}):

\begin{align}\begin{aligned}
\lim_{T \to 0 } \frac{C_V}{\frac{\pi^2}{3} k_B T } 
= \lim_{T \to 0}\Big( \frac{\partial n}{\partial \mu} \Big)_T
+ \Delta (\mu) ,
\label{linear-T-coefficient}
\end{aligned}\end{align}

\noindent where $\Delta (\mu) = 
\omega \sum_{l_1 l_2}
(l_2-l_1)
\nu_{l_1} (\mu) \nu_{l_2}^\prime (\mu)/(\sum_l \nu_l (\mu))$. The additional van-Hove singularities in the non-equilibrium DoS also manifest themselves as singularities in the non-equilibrium specific heat as illustrated in Fig.~\ref{fig-DOS-CV}(a) and \ref{fig-DOS-CV}(b) for the same square lattice tight-binding model of the previous section.

\medskip

\textit{\color{blue}Discussion.} We have demonstrated the existence of a non-equilibrium FFL steady state in Floquet bands that features a collection of FFSs enclosed inside each other. To realize these states in experiments essentially two criteria should be met: first the driving frequency should exceed thermal broadening $\hbar \Omega \gg k_BT$ \footnote{And an analogous criteria should be met for other quasiparticle broadening effects such as disorder and electron life-times.}, so that the multiple Floquet quasi-energy bands can be resolved. In addition, the size of the additional jumps of the Fermi Dirac Staircase occupation, which are the dimensionless numbers $| \varphi_{a,l} \big|^2$ in Eq.\eqref{CleanLimit-Periodic-DensityMatrix}, should be sizable. The first non-trivial jump scales as $| \varphi_{l=\pm 1} \big|^2 \sim (ev_F |{\bf E}|/\hbar \Omega^2)^2$, at small field amplitudes, therefore the second criterion is that the light intensity, $I=c \epsilon_0 |{\bf E}|^2/2$, is comparable to the intensity scale $I_0=\hbar \Omega^4 /(8 \pi \alpha v_F^2)$, with $\alpha \approx 1/137$ the fine structure constant. We believe that these criteria can be comfortably met in a variety of platforms, and, in fact, are likely met in several of those in which MIRO and SdH oscillations are seen to coexist \cite{shi2015shubnikov,dmitriev2012nonequilibrium}. For example for MIRO experiments \cite{dmitriev2012nonequilibrium} with a frequencies of $\Omega/2\pi = 10$GHz, and $v_F =2\times 10^5 $m/s, the intensity scale is just $I_0\approx 0.2$W/m$^2$, illustrating that low frequencies greatly help in reducing the required power. However, we believe there can be completely different conditions and material platforms for accessing the FFL regime. For example, for the experiments of Ref.\cite{wang2013observation} that studied Floquet-Bloch states in the surface of topological insulators with mid-infrared pulses of frequency $\hbar \Omega = 120$meV, it is estimated that $| \varphi_{l=\pm 1} \big|^2 \sim (ev_F |{\bf E}|/\hbar \Omega^2)^2 \sim 0.25$, and therefore meets the criteria. For the experiments of Ref.\cite{mciver2020light} realizing the light-induced anomalous Hall effect in graphene ($v_F=10^6$m/s) with a similar mid-infrared frequency, the intensity scale is $I_0\approx 4 \times 10^{12}$W/m$^2$, which is the same as their typical pulse peak intensity. Therefore, these type of experiments are well posed to prepare the FFL with pump pulses and investigate its subsequent decay.

\textit{\color{blue}Acknowledgements.} We are thankful to Qianhui Shi,  Michael Zudov and Ivan Dmitriev for valuable discussions. We acknowledge
support by the Deutsche Forschungsgemeinschaft (DFG) through research grant project number 518372354.

\bibliography{floquet-fermi-liquid}

\clearpage

\appendix

\renewcommand{\theequation}{\thesection-\arabic{equation}}
\renewcommand{\thefigure}{\thesection-\arabic{figure}}
\renewcommand{\thetable}{\thesection-\Roman{table}}

\onecolumngrid

\section*{Supplemental Material for ``The Floquet Fermi Liquid''}

\section{Density matrix of a general system Hamiltonian coupled to the featureless fermionic bath}
\label{appendixA1}
In Refs.~\cite{shi2023berry,matsyshyn2023fermi}, we focused on analyzing the case of a diagonal system Hamiltonian. In the current study, we broaden our scope by considering more general, non-diagonal system Hamiltonian.

Following Eqs.~(2) to (16) from Ref.~\cite{matsyshyn2023fermi}, we have the open-system Schr{\"o}dinger's equation for a generic system coupled to the featureless fermionic bath:
\begin{align}\begin{aligned}
i \partial_t \ket{ \psi_n^{(j)} (t) } 
& = \big[ H_S (t) - i \Gamma \big] \ket{ \psi_n^{(j)} (t) } 
 + \lambda e^{- i \varepsilon_j (t-t_0)} \ket{ \chi_{n} },
\label{S-OSE-2}
\end{aligned}\end{align}
where $\lambda$ is the tunneling amplitude onto the bath, and $\Gamma= \nu_0 \lambda^2/2$, with $\nu_0$ the density of states of the bath.
This equation is a non-Hermitian version of the Schr{\"o}dinger equation in which the system Hamiltonian is dressed by a constant imaginary part ``$-i \Gamma$'', accounting for the decay into the bath. And it crucially includes an inhomogeneous term (the one proportional to $\lambda$) which accounts for the bath feedback effect (see Refs.~\cite{shi2023berry,matsyshyn2023fermi} for details). The one-body density matrix projected onto the system at time $t$ is then given by
\begin{align}\begin{aligned}
\rho_S (t) = \sum_{n, j} f_0 (\varepsilon_j) \ket{\psi_n^{(j)} (t)} \bra{\psi_n^{(j)} (t)} ,
\label{S-System-Occupation-0}
\end{aligned}\end{align}

Let's denote the time evolution operator of the closed system by $U_S(t, t')$, so that it satisfies the following equation:
\begin{align}\begin{aligned}
i \partial_t U_S(t, t') = H_S(t) U_S(t, t').
\label{S-Us}
\end{aligned}\end{align}
Next, we define an auxiliary state $\ket{\tilde{\psi}_n^{(j)}(t)}$ from the state $\ket{\psi_n^{(j)}(t)}$ as follows:
\begin{align}\begin{aligned}
\ket{\psi_n^{(j)}(t)} \equiv e^{-\Gamma(t - t_0)} U_S(t, t') \ket{\tilde{\psi}_n^{(j)}(t)}.
\label{S-aux-psi}
\end{aligned}\end{align}
By substituting Eq.~(\ref{S-aux-psi}) into Eq.~(\ref{S-OSE-2}), we derive the open-system Schr\"{o}dinger equation for the auxiliary state $\ket{\tilde{\psi}_n^{(j)}(t)}$:
\begin{align}\begin{aligned}
i \frac{\partial}{\partial t} \ket{\tilde{\psi}_n^{(j)}(t)} = \lambda U_S(t', t) \ket{\chi_{n}}
e^{- i (\varepsilon_j + i \Gamma) (t - t_0)},
\label{S-OSE-3}
\end{aligned}\end{align}
where we used the identity $U_S(t, t') U_S(t', t) = U_S(t, t) = 1$.
The solution for $\ket{\tilde{\psi}_n^{(j)}(t)}$ is therefore:
\begin{align}\begin{aligned}
\ket{\tilde{\psi}_n^{(j)}(t)} =-i \lambda 
\int_{t_0}^t U_S (t', t'') \ket{\chi_{n}} e^{- i (\varepsilon_j + i \Gamma) (t'' - t_0)} d t'' .
\label{S-tilde-psi}
\end{aligned}\end{align}
Thus, the solution for the amplitude of the open system problem is
\begin{align}\begin{aligned}
\ket{\psi_n^{(j)}(t)} = -i \lambda e^{-\Gamma(t-t_0)} \int_{t_0}^t U_S(t, t') \ket{\chi_{n}} e^{- i (\varepsilon_j + i \Gamma) (t' - t_0)}  d t'
\label{S-OSE-solution}
\end{aligned}\end{align}
By inserting Eq.~(\ref{S-OSE-solution}) into Eq.~(\ref{S-System-Occupation-0}) and taking $t_0 \to -\infty$ to obtain the late time steady state, we achieve
the Eq.~(\ref{Full-DensityMatrix}) in the main text:
\begin{align}\begin{aligned}
\rho_S (t) = \Gamma 
\int_{-\infty}^{+\infty} \frac{ \text{d} \epsilon }{\pi}
f_0 (\epsilon) U_\Gamma (t, \epsilon) U_\Gamma^\dagger (t, \epsilon),
\quad
U_\Gamma (t, \epsilon) = \int_{-\infty}^{t} \text{d} t' 
e^{\Gamma (t' - t) - i \epsilon  t' } U_S (t, t') ,
\label{S-Full-DensityMatrix}
\end{aligned}\end{align}
where we utilized the property that the featureless fermionic bath has a constant density of state $\nu_B (\epsilon) = 2\pi \sum_{j} \delta (\epsilon - \varepsilon_j ) \equiv  \nu_0$, and used $\Gamma \equiv \lambda^2 \nu_0/ 2$.

We note that we have used an initial condition in writing Eq.~(\ref{S-tilde-psi}) so that $\ket{\psi_n^{(j)} (t_0)} =0$, namely that in a distant past all particles are located in the bath. 
This assumption is convenient because it eliminates the transient part of the solutions, but it is not strictly necessary. 
This is because Eq.~(\ref{S-OSE-2}) is a inhomogeneous equation for which one can always add any solution to the homogeneous equation (the one with $\lambda=0$) in order to satisfy any given initial condition. 
But such solutions of the homogeneous equation would all decay to zero as $t \to \infty$ for any finite $\Gamma>0$.
This can be viewed as a type of ``irreversible radiation'' of the information of the initial state of the system onto the bath, which eventually erases all such information leading to the unique late-time steady state from Eq.~(\ref{S-Full-DensityMatrix}).

\section{Density matrix of a Floquet system Hamiltonian coupled to the ideal fermionic bath}
\label{appendixA2}

We will now restrict to periodic Hamiltonians, $H(t)=H(t+T)$. We will consider the situation when the coupling to the bath is weak (as indicated by $\Gamma \to 0$), such that the bath operates as an ``ideal'' thermal bath. Our objective is to demonstrate that under these conditions, Eq.~(\ref{Full-DensityMatrix}) from the main text simplifies to Eq.~(\ref{CleanLimit-Periodic-DensityMatrix}).

First, we denote Floquet eigenstates as $\ket{\psi_a^F (t)}$ satisfying:
\begin{align}\begin{aligned}
i \partial_t \ket{\psi_a^F (t)} = H_S (t) \ket{\psi_a^F (t)} .
\end{aligned}\end{align}
From the Floquet's theorem, $\ket{\psi_a^F (t)}$ can be expressed in terms of its Floquet harmonics:
\begin{align}\label{lth-Harmonic}\begin{aligned}
\ket{\psi_a^F (t)} = \sum_l e^{-i(\epsilon_a^F + l \Omega) t} \ket{\varphi_{a,l}} ,
\quad
\Omega = 2\pi / T .
\end{aligned}\end{align}
Using this formulation, we can express the unitary evolution operator for the system, denoted as $U_S(t,t')$, as follows:
\begin{align}\begin{aligned}
U_S(t,t') = \sum_a \ket{\psi_a^F (t)} \bra{\psi_a^F (t')} = \sum_{a, l_1, l_2} e^{-i \epsilon_a^F (t-t')} e^{-i \Omega (l_1 t - l_2 t')} \ket{\varphi_{a,l_1}} \bra{\varphi_{a,l_2}} .
\label{S-U-system}
\end{aligned}\end{align}

Next, we substitute Eq.~(\ref{S-U-system}) into Equation~(\ref{S-Full-DensityMatrix}) and obtain:
\begin{align}\begin{aligned}
U_\Gamma (t,\epsilon) = \sum_{a, l_1, l_2} \ket{\varphi_{a,l_1}} \bra{\varphi_{a,l_2}}
\frac{e^{-i \Omega (l_1-l_2) t} e^{-i \epsilon t}}{i \epsilon_a^F + i \Omega l_2 -i \epsilon+\Gamma} .
\end{aligned}\end{align}
Upon integrating over $\epsilon$, we obtain:
\begin{align}\begin{aligned}
& \rho_S (t) =\sum_{a, b, l_1, l_2} \ket{\varphi_{a,l_1}} e^{-i \Omega l_1 t}
\bigg(\sum_{l_3, l_4} \braket{ \varphi_{a,l_3} | \varphi_{b,l_4} } e^{-i \Omega ( l_4-l_3 ) t} 
\\
& \quad \times \frac{\Gamma}{2 \Gamma-i (l_4 \Omega + \epsilon_b^F - l_3 \Omega - \epsilon_a^F )}[ f_+ (\epsilon_a^F + l_3 \Omega ) + f_- ( \epsilon_b^F + l_4 \Omega ) ] \bigg) e^{i \Omega l_2 t}
\bra{\varphi_{b,l_2}} ,
\end{aligned}\end{align}
where $f_+ (\epsilon) = [ f_- (\epsilon) ]^*$ and they are given by:
\begin{align}\begin{aligned}
f_\pm (\epsilon) = \frac{1}{2} \pm \frac{i}{\pi}\Psi^{(0)}
\Big(\frac{1}{2} \pm i\beta \frac{\epsilon \mp i \Gamma -\mu}{2\pi}\Big),
\label{fpm}
\end{aligned}\end{align}
with $\Psi^{(0)}$ the 0-th order Polygamma function (or the digamma function).

Assuming a weak coupling to the bath (i.e., $\Gamma \to 0$), only the terms that satisfy the equation $l_4 \Omega + \epsilon_b^F -l_3 \Omega - \epsilon_a^F = 0$ are retained. Provided that Floquet band crossings are avoided, the equality $l_4 \Omega + \epsilon_b^F -l_3 \Omega - \epsilon_a^F = 0$ imposes the conditions $a = b$ and $l_3=l_4$. Consequently, the density matrix simplifies to:
\begin{align}\begin{aligned}
\lim_{\Gamma \to 0} \rho_S (t) = \sum_a p_a \ket{\psi_a^F (t)} \bra{\psi_a^F (t)},
\quad
p_a = \sum_l \braket{\varphi_{a,l}|\varphi_{a,l}} f_0 (\epsilon_a^F + l \Omega) .
\label{S-CleanLimit-Periodic-DensityMatrix}
\end{aligned}\end{align}
where we used the limit
\begin{align}\begin{aligned}
\lim_{\Gamma \to 0}  \frac{1}{2} \Big[ f_+ (\epsilon) + f_- (\epsilon) \Big]  = f_0 (\epsilon) .
\label{S-f_Gamma_2}
\end{aligned}\end{align}
in obtaining Eq.~(\ref{S-CleanLimit-Periodic-DensityMatrix}).
From the above we can see that Eq.~(\ref{Full-DensityMatrix}) indeed simplifies to Eq.~(\ref{CleanLimit-Periodic-DensityMatrix}) in the main text.

\section{Floquet petrurbation theory}
\label{appendixB}
\subsection{General theory}
In this Section, we introduce the detailed scheme for how to perturbatively solve the Schrodinger equation in the presence of a periodic drive, where the state's dynamics is defined by (for instructive purposes in this part of the appendix, we restore the units):
\begin{align}\begin{aligned}
    i\hbar\partial_t \ket{\psi_a(t) } = H_S(t)\ket{\psi_a(t) }=(H_0 + V(t))\ket{\psi_a(t) }.\label{Bsr}
\end{aligned}\end{align}
Due to the time periodicity of the Hamiltonian $H_S(t)= H_S(t+T)$ we can use the following Floquet expansion for the states:
\begin{align}\begin{aligned}
    \ket{\psi_a(t)} =e^{-i\epsilon^F_a(t-t_0)/\hbar} \sum_{l'=-\infty}^\infty e^{-il'\Omega(t-t_0)}\sum_{b} c^{l'}_{a,b}\ket{\chi_b}\label{Nstate}
\end{aligned}\end{align}
where $\epsilon^F_a$ is the Floquet energy of the state and $\ket{\chi_b}$ are the solutions of the unperturbed problem $H_0\ket{\chi_b}=E^{(0)}_b\ket{\chi_b}$. Essentially, we expanded the r.h.s. of the Eq.(\ref{lth-Harmonic}) in the known basis $\ket{\varphi_{a,l}} = \sum_{b} c^{l}_{a,b}\ket{\chi_b}$. Since the Flouqet energy is not uniquely defined but up to a shift on $k\hbar\Omega$ where $k\in\mathbb{Z}$ [$k\Omega$ can be absorbed into the dummy index $l'$ in Eq.(\ref{Nstate})], without losing the generality, we fix this ambiguity or the Floquet gauge by setting $c^{l,0}_{a,b} = \delta_{l,0}\delta_{a,b}$ in the absence of the perturbation. Meaning that state $\ket{\psi_a(t)}$ is adiabatically connected to the unperturbed state $\ket{\chi_a(t)}$.

Next, we substitute Eq.(\ref{Nstate}) into Eq.(\ref{Bsr}) and project the equation onto mode $l$ obtaining: 
\begin{align}\begin{aligned}\label{EigenEq}
    (E^{(0)}_{a}+l\hbar\Omega-H_{0})\sum_{b} c^{l}_{a,b}\ket{\chi_b}=\int_0^T\frac{dt}{T}\sum_{l'=-\infty}^\infty(V(t)-\Delta_{a})e^{i(l-l')\Omega(t-t_0)}\sum_{b} c^{l'}_{a,b}\ket{\chi_b}.
\end{aligned}\end{align}
where $\Delta_a = \epsilon^F_a-E_a^{(0)} = \mathcal{O}(V^1)$ is a perturbation induced correction to the $a$-th state's energy. The form of the equation above is convinient since righthandside $V(t)-\Delta_{a}$ is at least of the order of the perturbation, while $E^{(0)}_{a}+l\hbar\Omega-H_{0}$ is unperturbative. Eq.(\ref{EigenEq}) is sufficient to determine all the Floque amplitudes and energies. 

By projecting Eq.(\ref{EigenEq}) on the state $\bra{\chi_a}$:
\begin{align}\begin{aligned}
    (\Delta_a+l\hbar\Omega) c^{n}_{a,a} = \bra{\chi_a}\int_0^T\frac{dt}{T}\sum_{l'=-\infty}^\infty V(t)e^{i(l-l')\Omega(t-t_0)}\sum_{b} c^{l'}_{a,b}\ket{\chi_b},
\end{aligned}\end{align}
we obtain the equation used to determine the $\Delta_a$. For $l \neq 0$ from projectiong the Eq.(\ref{EigenEq}) onto a state different from $a$, namely $\bra{\chi_c}\neq \bra{\chi_a}$ we obtain:  
\begin{align}\begin{aligned}
    c^{l}_{a,c}=\frac{1}{E^{(0)}_{a}-E^{(0)}_{c}+l\hbar\Omega}\bra{\chi_c}\int_0^T\frac{dt}{T}\sum_{l'=-\infty}^\infty(V(t)-\Delta_{a})e^{i(l-l')\Omega(t-t_0)}\sum_{b} c^{l'}_{a,b}\ket{\chi_b},
\end{aligned}\end{align}
which is correct for any $a,c$ assuming that $E^{(0)}_c-E^{(0)}_a\neq l\hbar\Omega$ for any $a,b$ and $l$.  The above is used to obtain coefficients $c^{l\neq 0}_{a,c}$. For $l=0$, the inverse operator $H-E_a^{(0)}$ is well defined on the space ortogonal to $\ket{\chi_a}$. Thus, using $P^\perp_a = \sum_{d\neq a}\ket{\chi_d}\bra{\chi_d}$, we can write:
\begin{align}\begin{aligned}
    c^{0}_{a,c}=\delta_{a,c}+\bra{
    \chi_c}\frac{P^\perp_a}{E^{(0)}_{a}-H_{0}}\int_0^T\frac{dt}{T}\sum_{l'=-\infty}^\infty(V(t)-\Delta_{N})e^{-il'\Omega(t-t_0)}\sum_{b} c^{l'}_{a,b}\ket{\chi_b},\label{B6}
\end{aligned}\end{align}
that determines remaining $c^{l=0}_{a,c}$ coeficinets. Note, $\delta_{a,c}$ was added as a solution of the homogeneous equation.

By using the perturbation expansions of states and energies in powers of their smallness:
\begin{align}\begin{aligned}
    &c^{l}_{a,c} = \delta_{l,0}\delta_{a,c}+c^{l,(1)}_{a,c} + c^{l,(2)}_{a,c} +\ldots ,\\
    &\Delta_a =  \Delta_a^{(1)}+ \Delta_a^{(2)} +\ldots, 
\end{aligned}\end{align}
where the superscript $(n)$ indicates the correction's order, we find perturbed states $\ket{\psi_a(t)}$. After these states are normalised, we find the Floque amplitudes using their definition in Eq.(\ref{CleanLimit-Periodic-DensityMatrix}) of the main text.

\subsection{Application to parabolic fermions}\label{C2}
We now apply our theory to parabolic fermions coupled simultaneously to a constant magnetic and periodic electric field. The corresponding Hamiltonian is $H({\bf k},t) = [\hbar{\bf k}-e{\bf A}_{0}({\bf r})-e{\bf A}(t)]^2/(2m)$, which using
\begin{equation}
    \boldsymbol{\pi}=\hbar{\bf k}-e{\bf A}_{0}({\bf r}),\qquad a = \frac{l}{\sqrt{2}\hbar}(\pi_x-i\pi_y),\qquad a^\dagger = \frac{l}{\sqrt{2}\hbar}(\pi_x+i\pi_y),\qquad [a,a^\dagger]=1,\qquad l = \sqrt{\frac{c\hbar}{eB}},
\end{equation}
we rewrite as:
\begin{align}\begin{aligned}
H({\bf k},t) = \hbar\omega_c a^\dagger a - a z_t^*-a^\dagger z_t + c(t),
\end{aligned}\end{align}
here $z_t = e\sqrt{{\hbar\omega_c}/{(2m)}}(A^x(t)-iA^y(t))$, $c(t) = {\hbar\omega_c}/{2}+e^2{{\bf A}(t)^2}/{2m}$, $\omega_c$ is the cyclotron frequency, $m$ is the electron mass. The solutions of the above equation, in the absence of the electric field, are the Landau level states. 

We consider $V(t)$ to be small. Note, the magnetic field is threated unperturbatively and is part of $H_0 = \hbar\omega_c a^\dagger a$. We absorb the time-dependant constant term of the Hamiltonian, $c(t)$, into a phase of the wave-function as follows $\ket{\psi_N(t)} = e^{-i C(t)}\ket{N(t)}$, where $C(t)=c_0+\int_{t_0}^t c(t^\prime) dt^\prime/\hbar$. Now, we can write:
\begin{align}\begin{aligned}
\label{pertSR}
    i\hbar\frac{d}{dt}\ket{N(t)}=( H_0 + V(t) )\ket{N(t)},\qquad V(t)=- a (z^*_+ e^{i\Omega t}+ z^*_-e^{-i\Omega t})-a^\dagger (z_+ e^{-i\Omega t}+ z_-e^{i\Omega t}).
\end{aligned}\end{align}
Next we apply the theory from Eq.(\ref{Bsr}-\ref{B6}) for $\ket{\chi_b} = \ket{M}^{(0)}$ to be unperturbed Landau states with $E_M^{(0)} = M\hbar\omega_c, M\in [0\ldots +\infty]$, where $\hbar\omega_c/2$ energy shift was absorbed to the phase $C(t)$. We find the first order coefficients to be:
\begin{align}\begin{aligned}
    &c^{1,(1)}_{N,N-1} =-\frac{z^*_-\sqrt{N}}{\hbar\omega_c+\hbar\Omega},  &&c^{1,(1)}_{N,N+1} =-\frac{z_+\sqrt{N+1}}{-\hbar\omega_c+\hbar\Omega}, \\
    &c^{-1,(1)}_{N,N-1}=-\frac{z^*_+\sqrt{N}}{\hbar\omega_c-\hbar\Omega},&&c^{-1,(1)}_{N,N+1}=-\frac{z_-\sqrt{N+1}}{-\hbar\omega_c-\hbar\Omega}, \label{B10}
\end{aligned}\end{align}
and $\Delta_N^{(1)} = 0$.
Note the states are yet to be normalised. All the coefficients beyong those appearing above are second or higher order in powers of electric field, which contribute as at least of third order correction to the amplitudes. The second order correction to the energy is found to be $\Delta_N^{(2)}=-|z_+|^2/({\hbar\omega_c-\hbar\Omega}) -{|z_-|^2}/({\hbar\omega_c+\hbar\Omega})$, where:
\begin{align}\begin{aligned}
|z_\pm|^2 = {\frac{\hbar\omega_c e^2}{2m\Omega^2}}(|{\bf E}|^2\pm i[{\bf E}\times{\bf E}^*]_z).\label{S-zpm}
\end{aligned}\end{align}

After the state normalization using Eqs.~(\ref{B10}) we find the the Floquet amplitudes up to the second order in powers of the electric field, which are given by:
\begin{align}\begin{aligned}
|\varphi_{N,\pm 1}|^2 =\frac{\epsilon_N^F}{\hbar\omega_c}R_+\pm \frac{R_-}{2}+\mathcal{O}(E^4) ,
\qquad
|\varphi_{N,0}|^2 = 1-2\frac{\epsilon_N^F}{\hbar\omega_c}R_++\mathcal{O}(E^4), \label{C12}
\end{aligned}\end{align}
and the Floquet energy of the N-th Landau level given by $\epsilon_N^F=(N+{1}/{2})\hbar\omega_c+\Delta E$, where
\begin{align}\begin{aligned}\label{Rpm}
R_\pm = \frac{|z_+|^2}{(\hbar\omega_c-\hbar\Omega)^2}\pm\frac{|z_-|^2}{(\hbar\omega_c+\hbar\Omega)^2},
\qquad
\Delta E = \frac{e^2|{\bf E}|^2}{m\Omega^2}-\frac{|z_+|^2}{\hbar\omega_c-\hbar\Omega} -\frac{|z_-|^2}{\hbar\omega_c+\hbar\Omega},
\end{aligned}\end{align}

 We can see that the oscillating electric field produces a uniform ($N$ independent) energy shift to all the Landau levels energies, which effectively redefines the chemical potential. Note, the dominant contribution to the oscillation of the $G_{\rm eff}$ comes from the terms in summation over $N$ (see Eq.(\ref{Geff})), when $\epsilon_N^F\approx\mu$, namely levels close to the Fermi Surface. For large chemical potentials, $\epsilon^F_N/\hbar\omega_c \gg 1$, the factor $R_-$ in Eq.(\ref{C12}) is negligible, yet it is interesting to note that $R_-$ is responsible for the imbalance of the occupation of the $l = 1$ vs $l = -1$ Floquet Fermi surfaces. Notice also that both $R_\pm$, according to Eq.(\ref{S-zpm}), are sensitive to the electric field polarization and, therefore, can be controlled by changing the degree of the polarization (e.g. linear vs circular) of the driving electric field. 

\section{Magnetic oscillations}
\label{appendixC}
In this Section, we show the detailed derivation of the system magnetic oscillations of several quantities, including the time-averaged system energy. We will employ the following Poisson summation formula:
\begin{align}\begin{aligned}
    \sum_{N=0}^\infty F(N+1/2) = \int_0^\infty F(x) dx +2 \sum_{k=1}^\infty \int_0^\infty (-1)^k F(x)\cos(2\pi kx )dx.\label{Poisson}
\end{aligned}\end{align}
Let us assume that the function $F(x)=f_0(x)g(x)$ can be represented as a multiplication of the $f_0$, Fermi-Dirac function, with a real function $g(x)$. Then the second integral of the r.h.s. of the Eq.(\ref{Poisson}) can be rewritten as:
\begin{align}\begin{aligned}
\int_0^\infty F(x)\cos(2\pi kx )dx 
= \int_0^\infty f_0(x) g(x)\cos(2\pi kx )dx 
= \int_0^\infty f_0(x) d\Big[\int_0^{x}g(y)\cos(2\pi ky )\Big]
\\
=-\int_0^\infty \frac{df_0(x)}{dx} \Big[\int_0^{x}g(y)\cos(2\pi ky )dy\Big] dx
\approx-{\rm Re}
\Big\{
\int^\infty_{\color{red}-\infty}  \frac{df_0(x)}{dx} \Big[\int_0^{x}g(y)e^{2\pi i ky }dy\Big] dx
\Big\},
\label{fcalc}
\end{aligned}\end{align}
where the surface term vanishes since the $f_0(\infty) \rightarrow 0$ and term at $x=0$ is zero due to the integral in the brakets. We replaced the lower limit of the integration to $-\infty$, which is a good approximation if $\mu/\hbar\omega_c\gg1$. In the last step, we also rewrote $\cos(2\pi kx)$ as an exponent since this form simplifies the integration over $x$. We aim to compute the magnetic oscillations up to the second order in powers of the electric field, also assuming $\mu \gg\hbar\omega_c$, $\mu \gg k_BT=\beta^{-1}$.

Interestingly, the expression in the square brackets in the last part of Eq.(\ref{fcalc}) is the zero-temperature result of the full expression if we set $x=\mu$ (note for $T=0$ $df_0[\epsilon]/d\epsilon\sim \delta(\epsilon-\mu)$). This is the reason why calculations for the magnetic oscillations are often carried out at zero temperature and, later on, weighted with the spectral weight of the Fermi-Dirac function:
\begin{align}\begin{aligned}
    \frac{\partial f_0}{\partial\epsilon}[\epsilon] = -\int_{-\infty}^\infty \frac{d\lambda}{2\pi} \frac{\pi\lambda/\beta}{\sinh(\pi\lambda/\beta)}e^{-i\lambda(\epsilon(x)-\mu)}=-\frac{\beta}{2}\frac{1}{1+\cosh(\beta(\epsilon-\mu))}.\label{fff}
\end{aligned}\end{align}

To be more concrete, let us assume $\epsilon(x) = \hbar\omega_c x+\Delta E$ and $g(y)$ to be a polynomial of $y$. Then the following shows the idea behind the further steps of the calculation:
\begin{multline}
    \int_0^\infty F(x)\cos(2\pi kx )dx \approx \hbar\omega_c{\rm Re}
\Big\{
\int^\infty_{-\infty}  \int_{-\infty}^\infty \frac{d\lambda}{2\pi} \frac{\pi\lambda/\beta}{\sinh(\pi\lambda/\beta)}e^{i\lambda(\mu-\Delta E)} e^{-i\lambda\hbar\omega_c x }\Big[\int_0^{x}g(y)e^{2\pi i ky }dy\Big]  dx
\Big\}=\\=\hbar\omega_c{\rm Re}
\Big\{
  \int_{-\infty}^\infty \frac{d\lambda}{2\pi} \frac{\pi\lambda/\beta}{\sinh(\pi\lambda/\beta)}e^{i\lambda(\mu-\Delta E)} \int^\infty_{-\infty}dx e^{-i\lambda\hbar\omega_c x }\Big[G(x)e^{2\pi i kx } + G_0\Big]
\Big\}\approx\\\approx {\rm Re}
\Big\{
  \int_{-\infty}^\infty d\lambda \frac{\pi\lambda/\beta}{\sinh(\pi\lambda/\beta)}e^{i\lambda(\mu-\Delta E)}  G\left(\frac{i\partial_\lambda}{\hbar\omega_c}\right)\delta\left(\frac{2\pi k}{\hbar\omega_c}-\lambda\right)
\Big\}\approx\\\approx 
  R_{T}(k)\underbrace{{\rm Re}
\Big\{e^{i\frac{2\pi k}{\hbar\omega_c}(\mu-\Delta E)}  G\left(\frac{\mu-\Delta E}{\hbar\omega_c}\right)\Big\}}_{\text{zero temperature result}},\label{details}
\end{multline}
where $\int_0^{x}g(y)e^{2\pi i ky }dy =G(x)e^{2\pi i kx } + G_0$. Polynomial $G(x)$ is to be determined for each quantity of interest, and in this sketch we discarded $G_0$ as a non-oscillating contribution. For the last step we performed integration by parts, and approximated $-i\partial_\lambda\approx \mu-\Delta E +\mathcal{O}(k_B T)$.
\subsection{Energy oscillations}

Let us now apply the above for the system time average energy calculation. The time average system energy up to the second order in EF can be found using Eq.(\ref{CleanLimit-Periodic-DensityMatrix}) of the main text and Eqs.(\ref{Nstate},\ref{B10}) as follows:
\begin{align}\begin{aligned}
\bar E(t) = 
N_\phi\int_0^T\frac{dt}{T}{\rm Tr}[\rho_S(t) H] 
= 
N_\phi\int_0^T\frac{dt}{T}
\sum_{N=0}^\infty 
p(\epsilon^F_N)\bra{\psi_N(t)}i\hbar d_t\ket{\psi_N(t)}
\\
=N_\phi\sum_{N = 0}^{\infty} 
\Big[{\sum_{l=-\infty}^\infty f_0 (\epsilon_N^F+l\hbar\Omega) |\varphi_{N,l}|^2} \Big]
\Big[\sum_{l^\prime=-\infty}^\infty[\epsilon_N^F+l^\prime\hbar\Omega]
|\varphi_{N,l^\prime}|^2\Big]
\label{bET}
\end{aligned}\end{align}
where we employed the Schrodinger equation $H(t)\ket{\psi_N(t)} = i\hbar d_t \ket{\psi_N(t)}$, and employed the normalised Floquet amplitudes as $|\varphi_{N,l}|^2 = \sum_{b} |c^{l}_{N,b}|^2/(\sum_{b,N} |c^{l}_{N,b}|^2)$, keeping terms up to the second order in powers of electric field. Using Eq.(\ref{C12}) and $\epsilon_N^F=(N+{1}/{2})\hbar\omega_c+\Delta E$ we can rewrite the r.h.s. of the Eq.(\ref{bET}) as a functions of the $N$-th sate Floqet energy:
\begin{align}\begin{aligned}
\sum_{l=-\infty}^\infty f_0(\epsilon_N^F+l\hbar\Omega) 
|\varphi_{N,l}|^2
=p(\epsilon^F_N) ,
\qquad
\sum_{l^\prime=-\infty}^\infty[\epsilon_N^F+l^\prime\hbar\Omega]|
\varphi_{N,l^\prime}|^2
=\varepsilon(\epsilon^F_N),
\end{aligned}\end{align}
where up to the second order in powers of electric field, these functions are (see Eq.(\ref{Rpm})):
\begin{align}\begin{aligned}\label{pofepsilon}
p(\epsilon) = 
f_0(\epsilon)
+
\frac{\epsilon}{\hbar\omega_c}R_+\Big[f_0(\epsilon+\hbar\Omega)+f_0(\epsilon-\hbar\Omega)-2f_0(\epsilon)\Big]
+
R_-\Big[f_0(\epsilon+\hbar\Omega)-f_0(\epsilon-\hbar\Omega)\Big]
\end{aligned}\end{align}
and $\varepsilon(\epsilon) = \epsilon + 2\hbar\Omega R_-$. Next, we employ the Poisson formula from Eq.(\ref{Poisson}) to get:
\begin{align}\begin{aligned}
    \bar{E}(B,\mu,T) = N_\phi\int_0^\infty dx p(\epsilon^F(x))\varepsilon(\epsilon^F(x))+2N_\phi \sum_{k=1}^\infty (-1)^k \int_0^\infty p(\epsilon^F(x))\varepsilon(\epsilon^F(x))\cos(2\pi kx )dx
\end{aligned}\end{align}
where $\epsilon^F(x) = x \hbar \omega_c+\Delta E$. Note, the form of the l.h.s. in Eq.(\ref{Poisson}) is evaluated at $x = N+1/2$. The first term in the equation above contains the Landau orbital diamagnetic effect, while the second is related to the magnetic oscillations, which are of our interest and we focus on the second term only. 

The integration of the magnetic oscillation term we perform using Eqs.(\ref{fcalc}-\ref{details}), while neglecting terms of order $k_B T$ and assuming $\mu/\hbar\omega_c\gg1$. Note that $p(\epsilon)$ in Eq.(\ref{pofepsilon}) is a sum of multiple terms of the form $f_0(x+a)g(x)$, thus according to Eqs.(\ref{fcalc}-\ref{details}) this already guarantees oscillations at multiple frequencies. After applying the procedure to each of them, we obtain the following result for the magnetic oscillations:
\begin{align}\begin{aligned}
\delta\bar{E}(B,\mu,T)-N_\phi\frac{\hbar\omega_c}{24}  
=N_\phi\sum_{l=\pm1,0}\sum_{k=1}^\infty 
\frac{\hbar\omega_c(-1)^k}{2k^2\pi^2} R_T(k)
\bigg(
\bigg[a_l+2b_l \frac{ \mu_{l}}{\hbar\omega_c}\bigg]
\cos\Big(2\pi k\frac{\mu_{l}}{\hbar\omega_c}\Big)
\\
+\bigg[2\pi k \Big(a_l +b_l \frac{ \mu_{l}}{\hbar\omega_c} \Big)
\frac{ \mu_{l}}{\hbar\omega_c}\bigg]\sin\Big(2\pi k\frac{\mu_{l}}{\hbar\omega_c}\Big)
\bigg),
\label{energyosc}
\end{aligned}\end{align}
where $a_1 = -a_{-1} = R_-, a_0 = 1$, $b_1 = b_{-1}= R_+, b_0 = -2R_+$ and $\mu_\eta = \mu-\eta\hbar\Omega -\Delta E $. The $R_T$ factor comes directly from the Fourier representation of the derivative of the Fermi-Dirac function in the Eq.(\ref{fff}), while the oscillations are the result of the integration in brackets of Eq.(\ref{fcalc}) over $y$. 

The part of the oscillations in Eq.(\ref{energyosc}) proportional to the $\sin$ arise because the bath does not conserve particle number whereas the part proportional to $\cos$ is the free energy oscillations discussed in the main text. One can also rewrite the oscillations in terms of the Fermi surface ratio to the magnetic field by employing:
\begin{align}\begin{aligned}
\mu = \frac{\hbar^2\pi k_F^2}{2\pi m} = \frac{\hbar^2S}{2\pi m}, 
\qquad 
\omega_c = \frac{eB}{m},\qquad 2\pi\frac{\mu}{\hbar\omega_c} =\frac{\hbar S}{eB},
\end{aligned}\end{align}
where $S$ is the area of the main Fermi surface and in the main text we adopted $e = \hbar = c =1$.
\subsection{Floquet free energy oscillations}
In this subsection, we provide some details on the derivation of the Floquet free energy oscillations, discussed in the main text. Using the definition Eq.(\ref{Geff}) of the main text, Eq.(\ref{Poisson}) and $|\varphi_{l,N}|^2 = |\varphi_{l}(\epsilon^F_N)|^2$, we rewrite the Floquet free energy as:
\begin{multline}
     -\frac{G}{N_\phi k_B T} =\sum_{l=-\infty}^\infty \int_0^\infty dx |\varphi_{l}(\epsilon^F({x}))|^2 \log\left[1 + e^{-\beta (\epsilon^F({x})+l\Omega-\mu)}\right]+\\+2 \sum_{l=-\infty}^\infty\sum_{k=1}^\infty (-1)^k \int_0^\infty |\varphi_{l}(\epsilon^F({x}))|^2 \log\left[1 + e^{-\beta (\epsilon^F({x})+l\Omega-\mu)}\right]\cos(2\pi kx )dx.
\end{multline}

For the next step, we will keep the generality of the derivation. We perform a similar protocol shown in Eq.(\ref{fcalc}), yet here we integrate by parts twice, obtaining:
\begin{equation}
     \frac{\delta G}{N_\phi k_B T} =2\beta\sum_{l=-\infty}^\infty \sum_{k=1}^\infty (-1)^k \int_0^\infty \frac{d f_0(\epsilon^F({x})+l\Omega)}{dx} \left\{\int_0^x dz \frac{d\epsilon^F({z}) }{dz}\left[\int_0^z|\varphi_{l}(\epsilon^F({y}))|^2\cos(2\pi ky )dy\right] \right\} dx.
\end{equation}
Next, we move to the perturbative consideration, which allows us to simplify the above, obtaining:
\begin{equation}
     \frac{\delta G}{N_\phi k_B T} \approx2\beta\hbar\omega_c\sum_{l=-1}^1 \sum_{k=1}^\infty (-1)^k \int_0^\infty \frac{d f_0(\epsilon^F({x})+l\Omega)}{dx} \left\{\int_0^x dz \left[\int_0^z|\varphi_{l}(\epsilon^F({y}))|^2\cos(2\pi ky )dy\right] \right\} dx,
\end{equation}
which by using $\int_0^z dx \int_0^x f(y)dy = \int_0^z (z-y)f(y)dy$ 
becomes:
\begin{equation}
     \frac{\delta G}{N_\phi k_B T} \approx2\beta\hbar\omega_c\sum_{l=-1}^1 \sum_{k=1}^\infty (-1)^k \int_0^\infty \frac{d f_0(\epsilon^F({x})+l\Omega)}{dx} \left[\int_0^x (x-y)|\varphi_{l}(\epsilon^F({y}))|^2\cos(2\pi ky )dy\right] dx.
\end{equation}
Finally, by discarding contributions of order $k_BT$, the oscillations are found as:
\begin{equation}
     \delta G\approx N_\phi \sum_{l =\pm1,0}\sum_{k=1}^\infty\frac{\hbar\omega_c}{2k^2\pi^2}(-1)^kR_T(k)\left(\delta_{l,0} + b_l \frac{\mu_l}{\hbar\omega_c}\right)\cos\left(2\pi k \frac{\mu_l}{\hbar\omega_c}\right),
\end{equation}
which is the result reported in the main text.
\subsection{Particle number oscillations}
In this subsection we provide some details on the derivation of the DoS oscillation. We start from the particle number calculation that by definition is:
\begin{align}\begin{aligned}
    n(B,\mu,T) =  \frac{1}{2 \pi l^2}\sum_{N = 0}^{\infty}p(\epsilon^F(N+1/2))= \frac{1}{2 \pi l^2}\int_0^\infty dx p(\epsilon^F(x))+ \frac{1}{ \pi l^2}\sum_{k=1}^\infty (-1)^k \int_0^\infty p(\epsilon^F(x))\cos(2\pi kx )dx
\end{aligned}\end{align}
and perform similar calculation as in the previous subsection obtaining the following oscillation part of the particle number:
\begin{align}\begin{aligned}
    \delta n(B,\mu,T) = \frac{1}{\pi l^2}\sum_{l=\pm1,0}\sum_{k=1}^\infty   \frac{(-1)^k}{4k^2\pi^2} R_T(k)\bigg[b_l\cos\left(2\pi k \frac{\mu_l}{\hbar\omega_c}\right) + 2\pi k \left( a_l + b_l\frac{\mu_\eta}{\hbar\omega_c}\right)\sin\left(2\pi k \frac{\mu_l}{\hbar\omega_c}\right)\bigg],
\end{aligned}\end{align}
which allows us to immediately find the oscillating part of DoS as:
\begin{align}\begin{aligned}
\delta\nu =
\lim_{T\rightarrow0} 
\frac{\partial \delta n}{\partial \mu}
\approx 
 \frac{1}{\pi l^2\hbar \omega_c}\sum_{l=\pm1,0}\sum_{k=1}^\infty   
(-1)^k R_T(k) 
\left( a_l + b_l\frac{\mu_l}{\hbar\omega_c}\right) 
\cos\Big(2\pi k \frac{\mu_l}{\hbar\omega_c}\Big)
\\
\approx \frac{2}{h \omega_c l^2 }\sum_{k=1}^\infty 
(-1)^k R_T(k)
\bigg[1-4R_+ \frac{\bar\mu}{\hbar\omega_c}
\sin^2\Big(\pi k\frac{\Omega }{\omega_c}\Big)\bigg]
\cos\Big(2\pi k \frac{\mu}{\hbar\omega_c}\Big),
\end{aligned}\end{align}
where we kept the leading term in $\mu_l/(\hbar\omega_c)$. Note, for the above summation to converge, one has to assume arbitrarily small, but non-zero temperature of the bath.

\section{Non-equilibrium specific heat at fixed particle number}
\label{appendixE}

In this section we aim to derive the non-equilibrium specific heat at a fixed particle number, by defining it as the derivative of the time-averaged energy of the system with respect to temperature. This derivation follows Eq.~(\ref{Density-of-states}) mentioned in the main text. 

First, we establish the relationship between the chemical potential $\mu$ and the temperature $T$, keeping the particle number constant. For the Floquet case, following Eq.~(\ref{Density-of-states}) in the main text, the total particle number is expressed as a function of both temperature and chemical potential:
\begin{align}\begin{aligned}
n_0 = 
\sum_{l=-\infty}^{+\infty} \int_{-\infty}^{+\infty} f_0(\epsilon) \nu_l (\epsilon) d \epsilon
= 
\sum_{l=-\infty}^{+\infty} \int_{-\infty}^\mu \nu_l (\epsilon) d \epsilon+\frac{\pi^2}{6}(k_B T)^2 \sum_{l=-\infty}^{+\infty} \nu_l^{\prime}(\mu)
+ O(T^4)
\label{n0-total}
\end{aligned}\end{align}
In the equation above, we have used the Sommerfeld expansion at low temperatures:
\begin{align}\begin{aligned}
\int_{-\infty}^{\infty} \frac{g(\varepsilon)}{e^{(\varepsilon-\mu) / k_B T }+1} 
d \varepsilon
=\int_{-\infty}^\mu g(\varepsilon) d \varepsilon
+
\frac{\pi^2}{6}(k_B T)^2 g^{\prime}(\mu)
+
O(T)^4
\end{aligned}\end{align}
By recasting $\mu = \mu_0 + \delta \mu$, and expanding the right-hand side of Eq.~(\ref{n0-total}) up to the first order of $\delta \mu$, we can deduce the relation between $\mu$ and $T$ at low temperatures:
\begin{align}\begin{aligned}
\mu \approx \mu_0 -\frac{\pi^2}{6}(k_B T)^2 \frac{\sum_{l=-\infty}^{+\infty} \nu_l^{\prime}(\mu_0)}{\sum_{l=-\infty}^{+\infty} \nu_l (\mu_0)}.
\label{mu-T-relation}
\end{aligned}\end{align}

Subsequently, we calculate the non-equilibrium specific heat at a fixed particle number. In the Floquet case, the averaged total energy is given by:
\begin{align}\begin{aligned}
\bar{E} (\mu, T) & =
\sum_{l=-\infty}^{+\infty} \int \frac{d {\bf k} }{(2 \pi)^d}
\big| \varphi_{{\bf k}, l} \big|^2
f_0 (\epsilon_{\bf k}^F + l \Omega)
\epsilon_{\bf k}^F 
=
\sum_{l=-\infty}^{+\infty} 
\int_{-\infty}^{+\infty}
d \epsilon
f_0(\epsilon) \nu_l (\epsilon) (\epsilon - l \Omega).
\end{aligned}\end{align}
By again using the Sommerfeld expansion at low temperature, we obtain
\begin{align}\begin{aligned}
\bar{E}(\mu, T)
=
\sum_{l=-\infty}^{+\infty}
\bigg[
\int_{-\infty}^\mu (\epsilon - l \Omega)\nu_l (\epsilon) d \epsilon
+
\frac{\pi^2}{6}(k_B T)^2
\Big(
\frac{\partial[ (\epsilon - l \Omega) \nu_l(\epsilon)]}{\partial \epsilon}
\Big)_{\epsilon=\mu}
+
O(T^4)
\bigg]
\end{aligned}\end{align}
We then make use of the chain rule
\begin{align}\begin{aligned}
C_V = \frac{\partial \bar{E}(\mu, T)}{\partial T} = \frac{\partial \bar{E}(\mu, T)}{\partial \mu} \frac{\partial \mu}{\partial T}+\frac{\partial \bar{E}(\mu, T)}{\partial T},
\end{aligned}\end{align}
then substituting $\partial \mu/\partial T$ from Eq.~(\ref{mu-T-relation}) into the equation above, we obtain the non-equilibrium specific heat, $C_V$, at fixed particle number as follows:
\begin{align}\begin{aligned}
C_V
& = \frac{\pi^2}{3} k_B T
\sum_{l =-\infty}^{+\infty} \nu_l (\mu_0)
+ \frac{\pi^2}{3} k_B T \frac{\Omega}{\sum_l \nu_l (\mu_0)} \sum_{l_2 l_1} (l_2-l_1)
\nu_{l_1} (\mu_0) \nu_{l_2}^\prime (\mu_0),
\label{S-CV}
\end{aligned}\end{align}
which is the Eq.~(\ref{linear-T-coefficient}) shown in the main text.

\section{Floquet amplitudes and van-Hove singularities in square lattice tight-binding Models}
\label{appendixF}

This section provides a discussion on the Floquet amplitude in the context of tight-binding models represented on a square lattice. 

\subsection{Harmonics of the periodic energy}

We begin by considering a tight-binding model on a square lattice with a lattice constant $a=1$, described by the dispersion relation
\begin{align}\begin{aligned}
\epsilon_{\bf k} 
= - 2 t \cos(k_x) - 2 t \cos(k_y) .
\label{S-square-lattice-model}
\end{aligned}\end{align}
We introduce a monochromatic AC driving expressed as
\begin{align}\begin{aligned}
{\bf A} (t) = [A_x \sin (\Omega t + \phi_x), A_y \sin (\Omega t + \phi_y) ].
\label{S-monochromatic-driving}
\end{aligned}\end{align}
As a result, the dispersion relation of the system evolves periodically as
\begin{align}\begin{aligned}
\epsilon_{\bf k}
\to 
\epsilon_{\bf k}(t)
=-2t \cos[k_x-A_x \sin (\Omega t+\phi_x)]-2t\cos[k_y-A_y \sin(\Omega t+\phi_y)].
\label{S-periodic-dispersion}
\end{aligned}\end{align}
Applying the Jacobi-Anger expansion allows us to derive the harmonics of the periodic energy as follows:
\begin{align}\begin{aligned}
\epsilon_{\bf k}^{(l)}
= \int_0^T \frac{d t}{T} \epsilon_{\bf k} (t) \exp (+i l \Omega t) 
& = 2t \Big[- \frac{e^{-i l(\phi_x+\pi)}}{2} J_{+l}(-A_x)  e^{+i k_x}-\frac{e^{-i l(\phi_x+\pi)}}{2} J_{+l}(+A_x)  e^{-i k_x}
\\
& -\frac{e^{-i l(\phi_y+\pi)}}{2} J_{+l}(-A_y)  e^{+i k_y}-\frac{e^{-i l(\phi_y+\pi)}}{2} J_{+l}(+A_y)  e^{-i k_y} \Big] .
\label{S-epsilon-l}
\end{aligned}\end{align}
Consequently, the Floquet energy can be calculated as
\begin{align}\begin{aligned}
\epsilon_{\bf k}^{(0)}
=-2t \cos (k_x) J_0(A_x)-2t \cos(k_y)J_0(A_y) .
\end{aligned}\end{align}
And we pay special attention to ${\bf k}$ points at ${\bf X} = (\pi, 0)$ and ${\bf Y} = (0, \pi)$ within the Brillouin zone, for which we obtain the following expressions
\begin{align}\begin{aligned}
& \epsilon_{\bf X}^{(l)} = \frac{1+(-1)^l}{2} \, [ +e^{-i l(\phi_x+\pi)} J_{+l}(+A_x)-e^{-i l(\phi_y+\pi)} J_{+l}(+A_y) ] ,
\\
& \epsilon_{\bf Y}^{(l)} = \frac{1+(-1)^l}{2} \, [-e^{-i l(\phi_x+\pi)} J_{+l}(+A_x)+e^{-i l(\phi_y+\pi)} J_{+l}(+A_y)] .
\end{aligned}\end{align}
From the above it follows that
\begin{align}\begin{aligned}
\epsilon_{\bf X}^{(l)}  
= - \epsilon_{\bf Y}^{(l)}
=
\begin{cases}
0 & l \text{ is odd}  \\
\text{non-zero} & l \text{ is even}
\end{cases}
\label{epsilon-even-odd}
\end{aligned}\end{align}

\subsection{Amplitudes of Floquet harmonics}

We now consider the amplitudes of Floquet harmonics, as expressed by the following equations:
\begin{align}\begin{aligned}
\varphi_{{\bf k}, l} & = \frac{1}{T}\int_{0}^{ T}  d t \bigg[ \exp ( + i l \Omega t  )
\times \exp \Big( - i \int_{0}^{t} d t' [\epsilon_{\bf k} (t') - \epsilon_{\bf k}^F  ] \Big) \bigg]
\\
& =\exp \bigg(
-\sum_{\substack{l_1=-\infty \\ l_1 \neq 0}}^{+\infty} 
\frac{\epsilon_{\bf k}^{(l_1)}}{l_1 \Omega}\bigg) \times \int_0^T \frac{d t}{T} \exp \bigg(
\sum_{\substack{l_1=-\infty \\ l_1 \neq 0}}^{+\infty} \frac{\epsilon_{\bf k}^{(l_1)} e^{-i l_1 \Omega t}}{l_1 \Omega}+i l \Omega t
\bigg) .
\label{S-varphi-l}
\end{aligned}\end{align}
Note that the phase factor $-\sum_{l_1 \neq 0} 
\epsilon_{\bf k}^{(l_1)} / (l_1 \Omega)$ is purely imaginary due to the fact that $\epsilon_{\bf k}^{(l_1)} = [\epsilon_{\bf k}^{(-l_1)}]^*$ which can be seen from Eq.~(\ref{S-epsilon-l}).
To compute $|\varphi_{\bf k}^{(l)}|^2$, we can disregard the global phase factor outside the integral and focus on the integral within it:
\begin{align}\begin{aligned}
\phi_{{\bf k}, l} = \int_0^T \frac{d t}{T} \exp \bigg(
\sum_{\substack{l_1=-\infty \\ l_1 \neq 0}}^{+\infty} \frac{\epsilon_{\bf k}^{(l_1)} e^{-i l_1 \Omega t}}{l_1 \Omega}+i l \Omega t
\bigg),
\quad
|\phi_{{\bf k}, l}|^2 = |\varphi_{{\bf k}, l}|^2.
\label{S-phi-l}
\end{aligned}\end{align}
We perform a change of variable, $z= \exp (i \Omega t)$:
\begin{align}\begin{aligned}
\phi_{{\bf k}, l}
& =\frac{1}{2 \pi i} \oint_{\arg [z]=0}^{\arg [z]=2 \pi} \prod_{\substack{l_1=-\infty \\ l_1 \neq 0}}^{+\infty} 
\exp \Big(\frac{\epsilon_{\bf k}^{(l_1)} z^{-l_1}}{l_1 \Omega} \Big) z^l \frac{d z}{z} ,
\end{aligned}\end{align}
We then express this equation by expanding the exponential functions using Taylor series and performing Cauchy's residue theorem:
\begin{align}\begin{aligned}
\phi_{{\bf k}, l}
=
\delta_{0, l} +\sum_{n_1 \times l_1=l} \frac{1}{n_{1} !}
\Big(\frac{\epsilon_{\bf k}^{(l_1)}}{l_1 \Omega}\Big)^{n_1}
+
\sum_{n_1 \times l_1+n_2 \times l_2=l} \frac{1}{n_{1} !}
\Big(\frac{\epsilon_{\bf k}^{(l_1)}}{l_1 \Omega}\Big)^{n_1} 
\frac{1}{n_{2} !}
\Big(\frac{\epsilon_{\bf k}^{(l_2)}}{l_2 \Omega}\Big)^{n_2} 
+\cdots .
\\
( l_{1,2,3, \cdots} \neq 0;
\quad
n_{1,2,3, \cdots} \geq 1 )
\label{S-phi_kl}
\end{aligned}\end{align}
Building upon equation~(\ref{S-phi_kl}), we deduce the following: if $l$ is odd, then
\begin{align}\begin{aligned}
n_1 \times l_1+n_2 \times l_2+n_3 \times l_3+\cdots \in \text { odd } 
\quad
\to
\quad
\text { at least odd numbers of }(n_i \in \text {odd}, l_i \in \text {odd}) \text { pairs }
\end{aligned}\end{align}
We then incorporate the conclusion from Eq.~(\ref{epsilon-even-odd}) which states that $\epsilon_{\bf X}^{(l)}  
= - \epsilon_{\bf Y}^{(l)} = 0$ when $l$ is odd, into Eq.~(\ref{S-phi_kl}). This yields:
\begin{align}\begin{aligned}
\phi_{{\bf X}, l}
= \phi_{{\bf Y}, l} = 0 ,
\qquad
l \in \text{odd} .
\label{S-phi-0}
\end{aligned}\end{align}
This result confirms that the contributions associated with the odd harmonics have vanishing weight at these special momenta, namely $\varphi_{(0,\pi), l} = \varphi_{(\pi,0), l} = 0$
for odd $l$.

\subsection{Floquet van-Hove singularities}

Let us first consider the DoS for equilibrium band without driving for the dispersion from
Eq.~(\ref{S-square-lattice-model}). For this model, the density of state is given by
\begin{align}\begin{aligned}
& \nu (\mu) = \frac{4}{|2t|} \int_0^{+\pi} \frac{d k_x}{2 \pi} \int_0^{+\pi} \frac{d k_y}{2 \pi} \delta(-\cos (k_x)-\cos (k_y)- \tilde{\mu} ),
\quad
\tilde{\mu} = \mu / 2t .
\end{aligned}\end{align}
By performing a change of variables, $u = - \cos (k_x)$ and $v = - \cos (k_y)$, we obtain
\begin{align}\begin{aligned}
& \nu(\mu) = 
\frac{2}{|t|}\frac{1}{(2 \pi)^2}
\int_{-1}^{+1} d u \int_{-1}^{+1} d v \frac{1}{\sqrt{1-u^2} \sqrt{1-v^2}} \delta(u+v-\mu) .
\end{aligned}\end{align}
Notably, the singularity in the above equation originates from the singularities of the integrand at 
\begin{align}\begin{aligned}
& u=+1, v=-1 \text 
{ or } u=-1, v=+1
\end{aligned}\end{align}
or equivalently, from the following special $\bf k$ points:
\begin{align}\begin{aligned}
& {\bf X} = (\pi, 0) \text { or } {\bf Y} = (0, \pi),
\end{aligned}\end{align}
leading to divergence of the integral at $\tilde{\mu}=u+v=0$ at half filling.

\medskip

We now proceed to introduce the driving as per Eqs.~(\ref{S-monochromatic-driving}) and (\ref{S-periodic-dispersion}). Following the definitions given in Eqs.~(\ref{clean_p_k}) and (\ref{Density-of-states}) in the main text, we obtain the following relationships:
\begin{align}\begin{aligned}
& \nu (\mu) = \sum_{l=-\infty}^{+\infty} \nu_l(\mu_l), 
\quad
\mu_l = \mu - l \Omega,
\end{aligned}\end{align}
and
\begin{align}\begin{aligned}
& \nu_l(\mu_l) = 
\int_{-\pi}^{+\pi} \frac{d k_x}{2 \pi} \int_{-\pi}^{+\pi} \frac{d k_y}{2 \pi} |\phi_{\bf k}^{(l)}|^2 \delta(\epsilon_{\bf k}^{(0)}-\mu_l)
=\sum_{\eta_x, \eta_y = \pm 1} \nu_l^{(\eta_x, \eta_y)} (\mu_l).
\end{aligned}\end{align}
Here, $\nu_l^{(\eta_x, \eta_y)} (\mu_l)$ correspond to different patches of the Brillouin zone, which we define as follows:
\begin{align}\begin{aligned}
& \nu_l^{(-1, -1)} (\mu_l) = \int_{-\pi}^{0} \frac{d k_x}{2 \pi} \int_{-\pi}^{0} \frac{d k_y}{2 \pi} |\phi_{\bf k}^{(l)}|^2 \delta(\epsilon_{\bf k}^{(0)}-\mu_l) ,
\quad
\nu_l^{(-1, +1)} (\mu_l) = \int_{-\pi}^{0} \frac{d k_x}{2 \pi} \int_{0}^{+\pi} \frac{d k_y}{2 \pi} |\phi_{\bf k}^{(l)}|^2 \delta(\epsilon_{\bf k}^{(0)}-\mu_l), 
\\
& \nu_l^{(+1, -1)} (\mu_l) = \int_{0}^{+\pi} \frac{d k_x}{2 \pi} \int_{-\pi}^{0} \frac{d k_y}{2 \pi} |\phi_{\bf k}^{(l)}|^2 \delta(\epsilon_{\bf k}^{(0)}-\mu_l) ,
\quad
\nu_l^{(+1, +1)} (\mu_l) = \int_{0}^{+\pi} \frac{d k_x}{2 \pi} \int_{0}^{+\pi} \frac{d k_y}{2 \pi} |\phi_{\bf k}^{(l)}|^2 \delta(\epsilon_{\bf k}^{(0)}-\mu_l),
\end{aligned}\end{align}
where $\eta_{x,y}=\pm 1$ indicates four different patches of the Brillouin zone. This division allows for a one-to-one change of variables given by
\begin{align}\begin{aligned}
u = - \cos(k_x) J_0 (A_x), 
\qquad
v = - \cos(k_y) J_0 (A_y) ,
\end{aligned}\end{align}
which subsequently results in Jacobians on different patches as shown below:
\begin{align}\begin{aligned}
\Big| \frac{\partial(k_x, k_y)}{\partial(u,v)} \Big|^{(\eta_x, \eta_y)}
= \frac{|\eta_x \eta_y|}{\sqrt{J_0^2 (A_x) -u^2} \sqrt{J_0^2 (A_y) -v^2} },
\quad
\eta_{x,y} = \pm 1 .
\end{aligned}\end{align}
By incorporating these transformations, we can express $\nu_l^{(\eta_x , \eta_y)}(\mu_l)$ as
\begin{align}\begin{aligned}
\nu_l^{(\eta_x , \eta_y)}(\mu_l) & =
\frac{2}{|t|}\frac{1}{(2 \pi)^2} \int_{-|J_0(A_x)|}^{+|J_0(A_x)|} d u \int_{-|J_0(A_y)|}^{+|J_0(A_y)|} d v \frac{|\phi_l^{\eta_x, \eta_y}(u, v)|^2}{\sqrt{J_0^2(A_x)-u^2} \sqrt{J_0^2(A_y)-v^2}} \delta(u+v-\tilde{\mu}_l) .
\label{S-nu-l}
\end{aligned}\end{align}
Here, $\phi_l^{\eta_x, \eta_y} (u, v)$ represents the transformation from $\phi_{{\bf k}, l}$ on each of the four patches. This means that we need to choose different signs for $\sin (k_x)$ and $\sin (k_y)$ based on the values of $\eta_x$ and $\eta_y$.

Examining Eq.~(\ref{S-nu-l}), it becomes evident that, unlike in the non-driven case, each of the $\nu_l (\mu_l)$ can potentially have two separate van-Hove singularities that could originate from the singularities of the integrand at
\begin{align}\begin{aligned}
& u=+J_0(A_x), v=-J_0(A_y) 
\text { or } 
u=-J_0(A_x), v=+J_0(A_y),
\end{aligned}\end{align}
or, equivalently, from the special ${\bf k}$ points
\begin{align}\begin{aligned}
& {\bf X} = (\pi, 0) 
\text { or } 
{\bf Y} = (0, \pi),
\end{aligned}\end{align}
which create van-Hove singularities when the conditions
\begin{align}\begin{aligned}
\tilde{\mu}_l = +J_0(A_x)-J_0(A_y) 
\text { or } 
\tilde{\mu}_l = -J_0(A_x)+J_0(A_y)
\end{aligned}\end{align}
are met, in which $\tilde{\mu}_l = (\mu - l \Omega)/(2t)$.
Moreover, we also observe that: 

\medskip

(i) when $A_x = A_y$, the two distinct singularities merge into a single singularity;

\medskip

(ii) as demonstrated in Eq.~(\ref{S-phi-0}), for odd values of $l$, the conditions $\phi_{{\bf X}, l} = \phi_{{\bf Y}, l} = 0$ and therefore $|\varphi_{{\bf X}, l}|^2 = |\varphi_{{\bf Y}, l}|^2 = 0$ hold, which prevent the appearance of van-Hove singularities for $\nu_l (\mu_l)$ when $l$ is odd in the current model.;

\medskip

(iii) Equations (\ref{S-epsilon-l}), (\ref{S-varphi-l}), and (\ref{S-nu-l}), with particular emphasis on Eq.(\ref{S-phi-l}), offer a non-perturbative expression for numerically determining the density of states. The convergence stems from the relationship $\epsilon_{\bf k}^{(l)} \sim J_l (A_{x,y})$, which decays rapidly as $l$ increases for a given $A_{x,y}$. For our numerical evaluations concerning Eq.(\ref{S-phi-l}), we sum over values of $l_1$ ranging from $-100$ to $+100$, ensuring well-converged results.
\end{document}